\documentclass[11pt,twocolumn,twoside,a4paper,amsmath,amssymb,aps,showkeys,showpacs]{revtex4}

\usepackage{amsfonts}
\usepackage{graphics} 
\usepackage{epsfig}
\usepackage{fancyheadings}

\begin{document}
\thispagestyle{myheadings}
\rhead[]{}
\lhead[]{}
\chead[Abramovsky V.A.]{Are multiple parton interactions important at high energies?}

\title{Are multiple parton interactions important at high energies? New types of hadrons production processes}

\author{Abramovsky V.A.}

\email{Victor.Abramovsky@novsu.ru}

\affiliation{Novgorod State University, Novgorog-the-Great,
Russia}

\begin{abstract}
Hadrons interaction at high energies is carried out by one color
gluon exchange. All quarks and gluons contained in colliding
hadrons take part in interaction and production of particles. The
contribution of multiple parton interactions is negligible.
Multiple hadrons production at high energies occurs only in three
types of processes. The first process is hadrons production in
gluon string, the second is hadrons production in two quark
strings and the third is hadrons production in three quark
strings. In proton-proton interaction production of only gluon
string and two quark strings is possible. In proton-antiproton
interaction production of gluon string, two quark strings and
three quark strings is possible. Therefore multiplicity
distributions in proton-proton and proton-antiproton interactions
are different.\end{abstract}

\pacs{12.40Nn, 13.85.Hd, 13.85Lg}

\keywords{multiplicity distribution, proton, antiproton, quark
string, gluon string}

\thanks{Author is very grateful to RFBR for financial support, grant 09-02-08428-z, and  Organizing Committee for hospitality}

\maketitle


\section{Introduction}
We will consider possible types of processes (or, more accurately,
subprocesses) of hadrons production that give contribution to
multiple production. The main picture of hadrons interaction at
high energies in the last 40 years is picture of multiple
interactions. It is based on Gribov reggeon theory~\cite{bib1} in
which hadrons interaction is described by multiple reggeon
exchange between components of hadrons. Many experimental data
were described in this theory. It is important for us that
absorbtive parts of every reggeon diagram correspond to definite
type of inelastic process, namely to multiple pomeron showers.
These subprocesses widen multiplicity distribution function of
secondary hadrons $P_n=\sigma_n/\sigma_{tot}$ and in idealized
case at asymptotic high energies lead to oscillations~\cite{bib2}.
This behavior is characteristic to any models of multiple
scattering no matter how they are named or what hadrons
constituents are discussed.

The low Constituents Number Model (LCNM) was proposed in
1980~\cite{bib3}. Introduction of this model was caused by the
fact that it is impossible to adjust low reduction of elastic
scattering diffractive cone with energy increase and large mean
multiplicity in models, based on cascades of particles both
colorless and quarks and gluons.

\section{LCNM and quasi-eikonal approach}
In the LCNM gluons density in rapidity space is low in wave
function of initial state and real hadrons are produced by decay
of tubes (strings) of color field. Interaction results from gluon
exchange, hadrons gain color charge. Color field tube is formed
between flying hadrons and than it decays to secondary hadrons.
Since transverse gluons have sense only in region of weak
connection $\alpha_s(k_{g\perp}^2)\ll 1$ then values of transverse
momenta of these gluons are large, $k_{g\perp}\sim2$~GeV. So
probability of gluon appearance in spectrum of projectiles is low
and thus smallness of $\alpha'_p$ is explained.

It was shown in~\cite{bib4} that total cross sections of $pp$ and
$p\bar{p}$ are well described in LCNM by contributions of initial
state configurations corresponding to only valent quarks, valent
quarks with one or two additional gluons.
\begin{eqnarray}
\label{1} \sigma_{tot}^{p(\bar{p})p}=63.52s^{-0.358}\mp
35.43s^{-0.56}+\nonumber\\+\sigma_0+ \sigma_1\ln s+\sigma_2(\ln
s)^2
\end{eqnarray}
The values of parameters are $\sigma_0=20.08\pm0.42$,
$\sigma_1=1.14\pm0.13$, $\sigma_2=0.16\pm0.01$.

Model parametrization of multiple exchange is quasi-eikonal model
in which formula of total cross section can be written as

\begin{equation}
\label{2}\displaystyle\sigma_{tot}=\frac{8\pi(\alpha'_p\ln
s+R^2)}{c}\, F\left(\frac{z}{2}\right),
\end{equation}
where $F(z)=\sum_{m=1}^\infty(-1)^{m-1}z^m/(m\cdot m!)$,
$z/2=g^2s^\Delta c/(8\pi(\alpha'_p\ln s+R^2))$.

We place parameters corresponding to the following
sets~\cite{bib4a}
$$\begin{array}{llll}
\mbox{I}.&\Delta=\alpha_p(0)-1=0.21,&g^2/8\pi=1.5\;\mbox{GeV}^{-2};\\[3mm]
\mbox{II}.&\Delta=\alpha_p(0)=0.12,&g^2/8\pi=2.4\;\mbox{GeV}^{-2};\\[3mm]
\mbox{III}.&\Delta=\alpha_p(0)=0.07,&g^2/8\pi=3.64\;\mbox{GeV}^{-2}.\\
\end{array}
$$
and $c=1.5$, $\alpha'_p=0.25\;\mbox{GeV}^{-2}$,
$R^2=3.56\;\mbox{GeV}^{-2}$. These corresponding graphs are
presented in Fig.~1, data were taken from~\cite{bib5}.

\begin{figure}
\includegraphics[scale=0.43]{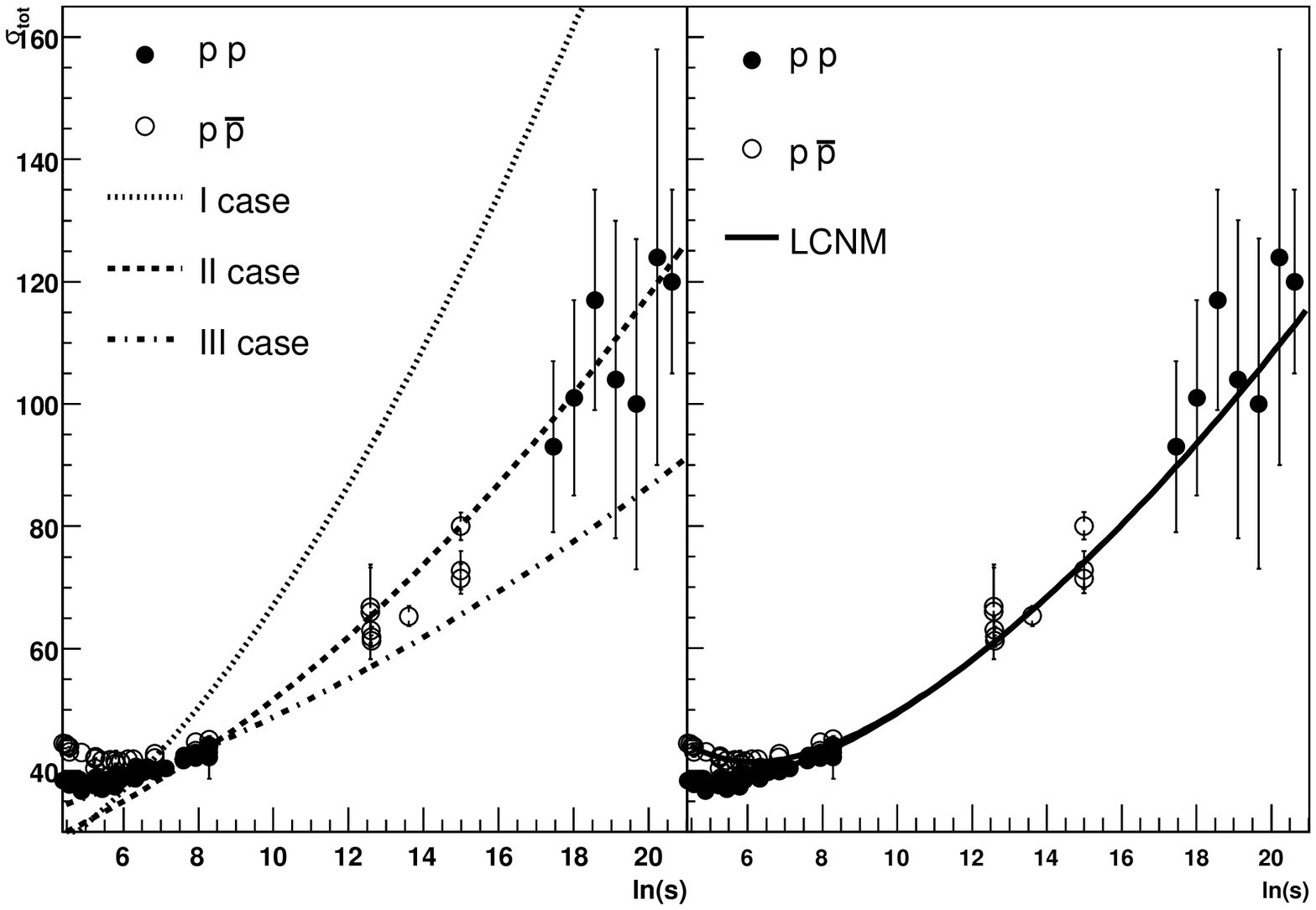}
\caption{Fitting of $\sigma_{tot}$ of $pp$ and $p\bar{p}$
collisions in quasi-eikonal model and LCNM}
\end{figure}

The differential cross section of elastic scattering at high
energies ($\sqrt{s}\geqslant53$~GeV) in LCNM is described as
following
\begin{eqnarray}\label{3}
\frac{d\sigma^{el}}{dt}=\frac{1}{16\pi}\left[\sigma_0+\sigma_1\ln
s+\sigma_2(\ln
s)^2\right]^2\times\nonumber\\\times\left(1+\rho^2\right)\exp\{-B(s)|t|\}
\end{eqnarray}
Experimental data at $\sqrt{s}=53$, 62, 546, 1800~GeV~\cite{bib6}
were fitted using this formulae. Only parameter $B(s)$ was varied.

In quasi-eikonal model $d\sigma^{el}/dt$ is defined as follows
\begin{eqnarray}\label{4}
\displaystyle\frac{d\sigma^{el}}{dt}=\frac{4\pi\left(\alpha'_p\ln
s+R^2\right)^2}{c^2}\left[\sum_{m=1}^\infty\frac{(-1)^{m-1}}{m\cdot
m!}\left(\frac{z}{2}\right)^m \times\right.\nonumber\\
\displaystyle\times\left.\exp\left\{-\frac{(\alpha'_p\ln
s+R^2)}{m}\,|\,t|\right\}\right]^2
\end{eqnarray}
We took parameters from II set which give good description of
total cross sections. The comparison of fits is given in Fig.~2
and 3.

\begin{figure}
\includegraphics[scale=0.43]{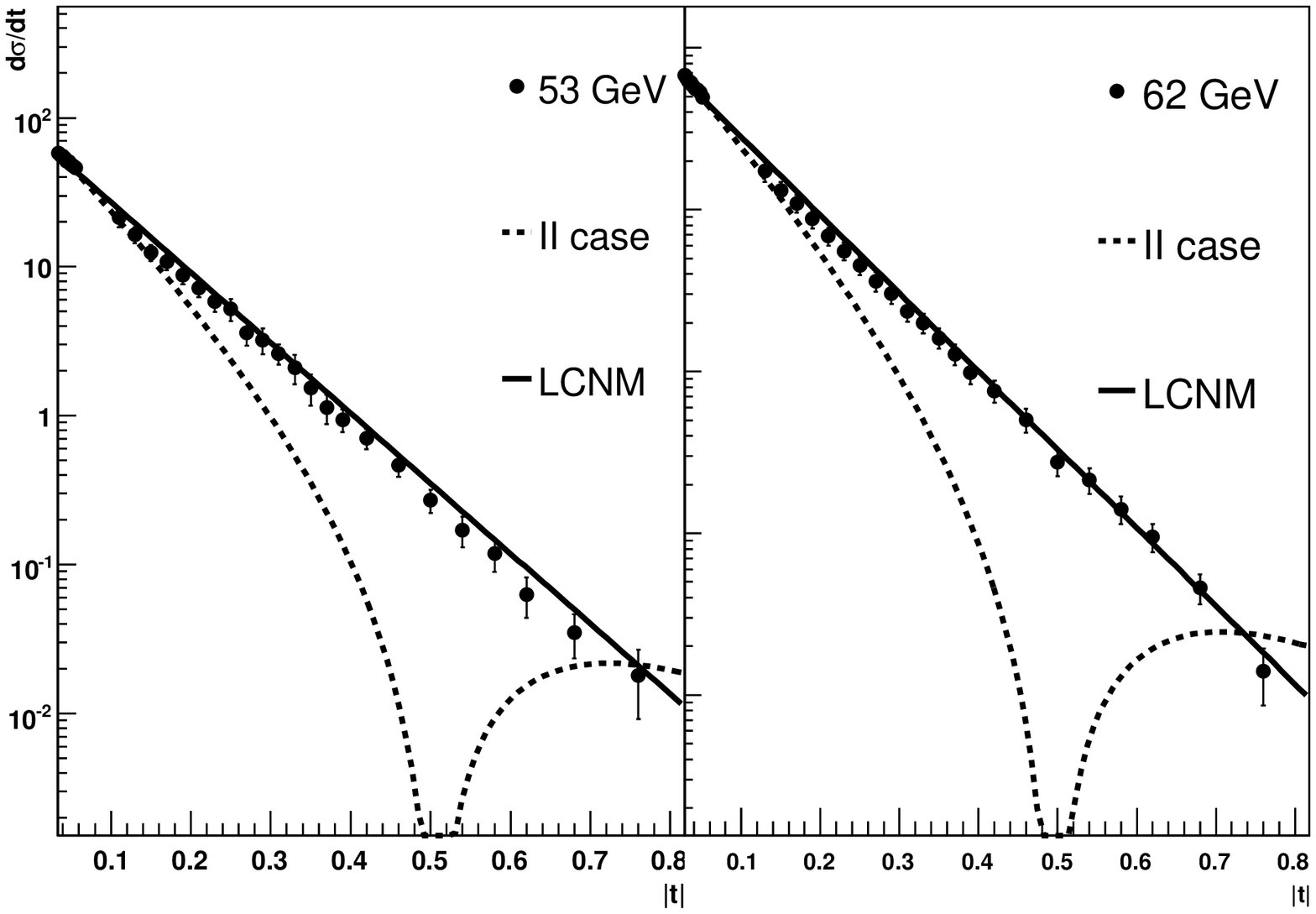}
\caption{Fitting of $d\sigma^{el}/dt$ of $pp$ and $p\bar{p}$
collisions in quasi-eikonal model and LCNM}
\end{figure}
\begin{figure}
\includegraphics[scale=0.43]{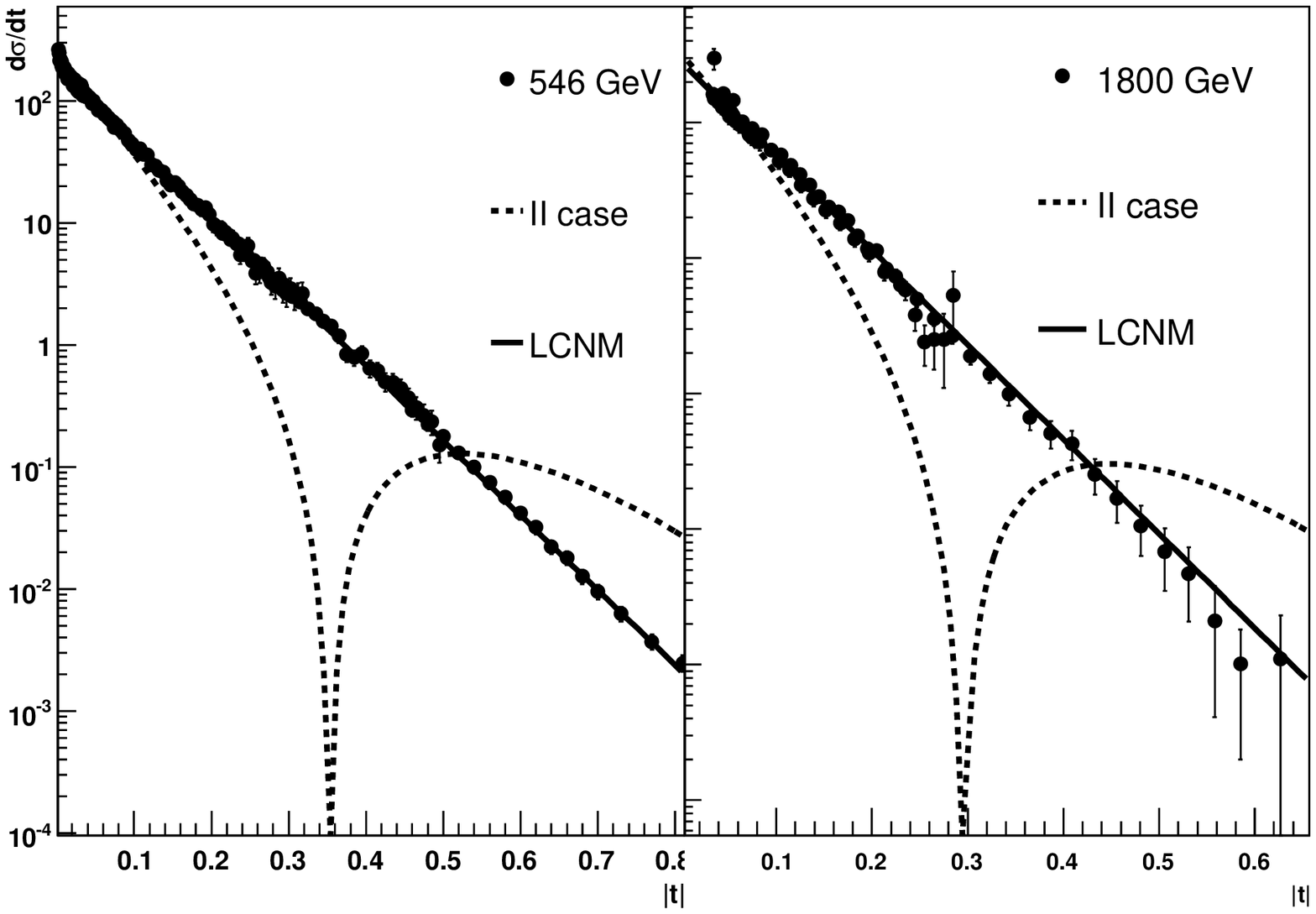}
\caption{Fitting of $d\sigma^{el}/dt$ of $pp$ and $p\bar{p}$
collisions in quasi-eikonal model and LCNM}
\end{figure}

Values of $B(s)$ obtained from fitting of $d\sigma^{el}/dt$ were
then fitted by quadratic function with the following parameters
$$
B(s)=7.12+0.34\ln s+0.02\ln^2 s.
$$

Elastic cross sections at high energies are described by the
following formulaes in LCNM~(\ref{5}) and quasi-eikonal~(\ref{6}).

\begin{equation}
\label{5}
\sigma^{el}=\frac{1}{16\pi}\frac{\left(\sigma_0+\sigma_1\ln
s+\sigma_2\ln^2s\right)^2}{B(s)}
\end{equation}

\begin{equation}
\label{6} \sigma^{el}=\frac{4\pi(\alpha'_p\ln
s+R^2)}{c^2}\left[2F\left(\frac{z}{2}\right)-F(z)\right]
\end{equation}
Graphs are shown in Fig.~4. Thus we can state that LCNM gives
better description of experimental data than the best set of
parameters in quasi-eikonal approach.

\begin{figure}
\includegraphics[scale=0.43]{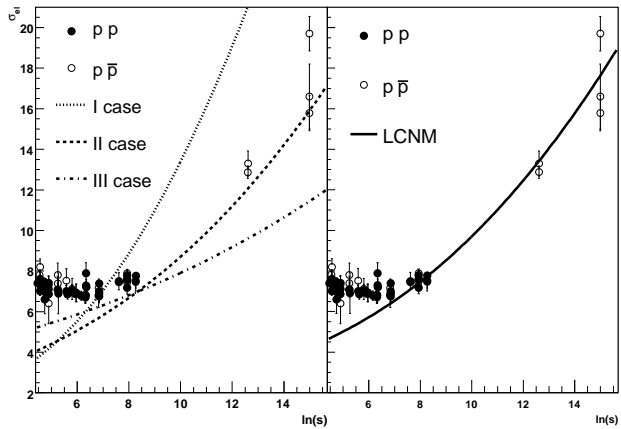}
\caption{Fitting of $\sigma^{el}$ of $pp$ and $p\bar{p}$
collisions in quasi-eikonal model and LCNM}
\end{figure}

\section{Gluon string}
There are no other characteristic sizes for configuration with
only valence quarks but hadrons sizes. Value of chromodynamics
constant $\alpha_s$ is large at these sizes. After color gluon
exchange two objects with gluon charges move apart, their sizes
coincide with colliding hadrons sizes. Large number of gluons is
produced in gluon string because of running constant $\alpha_s$
large value. This process can be shown in diagrams like the one
shown in Fig.~5. Gluon pairs production continues until pair
masses approaches hadrons masses. There is no suppression on
energy because exchange is of vector type. All these diagrams have
the same order of magnitude in given order of coupling constant
and every one of them corresponds to definite final hadrons state.
Number of these diagrams is infinite in principle. Hence secondary
hadrons multiplicity in final state as random variable has to obey
normal distribution because of central limit theorem of
probability theory.

\begin{figure}
\includegraphics[scale=0.6]{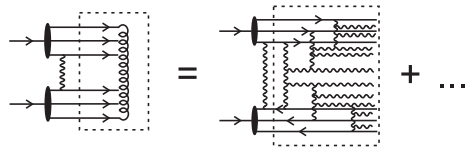}
\caption{Hadrons production in gluon string decay}
\end{figure}

\section{Quark string}
In configuration with one and two transverse gluons scale is
defined by large transverse momentum $k_g^\perp\simeq 2$~GeV and
running constant $\alpha_s$ is low. Quarks in moving apart hadrons
are greatly separated and quark strings are produced.
Quark-antiquark pairs produced in string field are virtual and
they come out on mass shell by tunneling. Momenta of these quarks
appearing at mass shell are equal to zero in center-of-mass system
of moving apart quarks. The break-up of string in models of Lund
type is schematicly described by diagram in Fig.~6a. Let us
associate this diagram to diagram in Fig.~6b. Diagram is Lorenz
invariant, that is in any reference system string break-up begins
from slow quarks. Color and spin correlations are essential. So
secondary hadrons multiplicity distribution will vary from Poisson
and normal distributions.

\begin{figure*}
\includegraphics[scale=0.6]{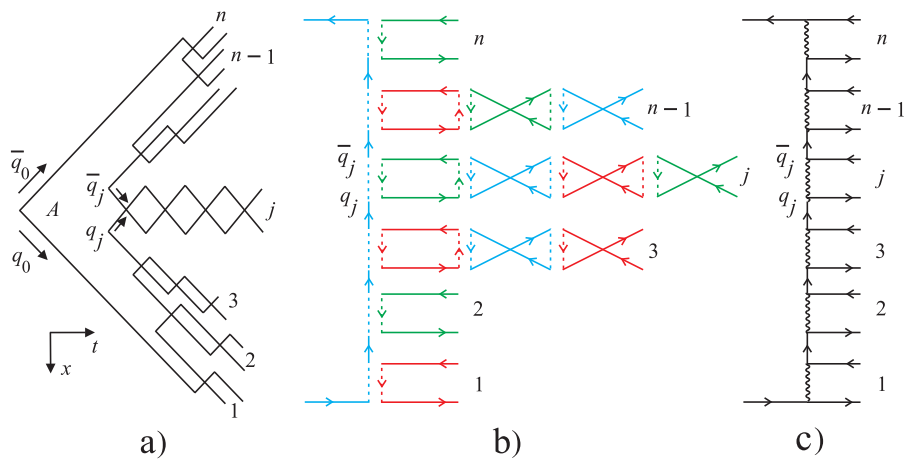}
\caption{Hadrons production in quark string decay}
\end{figure*}

Transverse sizes of string arising in configurations with one and
two gluons are small because they have to consume these gluons.

Let us consider $pp$ scattering. One additional gluon is consumed
by quark string arising between quark and diquark thereby the
second quark string is allocated. Corresponding inelastic process
is hadrons production in two independent quark strings. Since in
$pp$ interaction quark strings production is possible only between
quark and diquark, configuration with two additional quarks leads
to the same result.

In $p\bar{p}$ interaction configuration with one gluon leads to
hadrons production in two quark strings. Also there is the same
process of hadrons production in two quark strings when two gluons
are consumed by the same quark string. However, if two gluons are
consumed by different quark strings, they allocate the third quark
string thereby. So one more type of inelastic process is possible
in $p\bar{p}$ interaction, it is hadrons production in three quark
strings.

We would like to stress out that there are no other inelastic
processes. There have to be four additional gluons in initial
state in order to produce two pomeron showers. Probability of this
is negligible.

In this way there are three types of inelastic processes at high
energies: 1) hadrons production in gluon string; 2) hadrons
production in two quark strings; 3) hadrons production in three
quark strings. There are no other processes in our opinion.

\section{Single diffraction cross section}
\begin{figure*}
\includegraphics[scale=0.55]{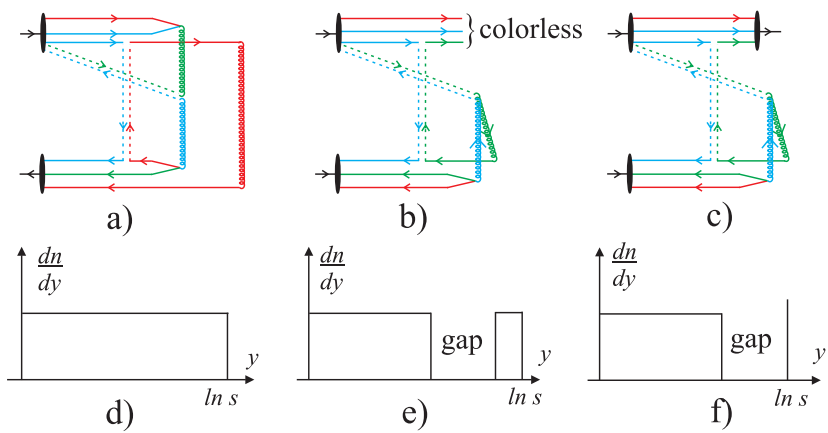}
\caption{Color diagrams and rapidity spectrums}
\end{figure*}
\begin{figure}
\includegraphics[scale=0.43]{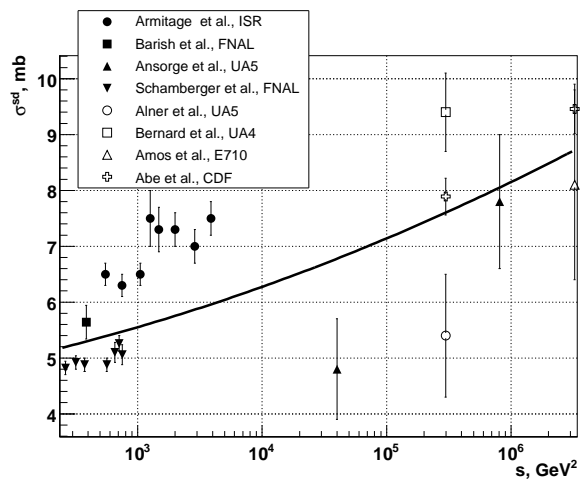}
\caption{Fitting of $\sigma^{sd}$ of $pp$ and $p\bar{p}$
collisions in  LCNM}
\end{figure}

In LCNM single diffraction to large masses can be described as
follows. From all possible gluon exchanges on Fig.~7a we will
choose the one that leads to colorless configuration from three
leading quarks. Color string arises between system of three
antiquarks and additional gluon. It is double quark string.

Diagrams in  Fig.~7a and 7b correspond to spectrums $dn/dy$, $y$
means rapidity, in figures Fig.~7d and 7e. In Fig.~7e beam with
large mass $M$ and beam of small mass $\mu$ from diffractive
proton are shown. If three quarks system does not diffract then
leading proton arises, which is shown in Fig.~7c, spectrum is
shown in Fig.~7f.

Value of diffractive gap in rapidity space is defined by energy of
additional gluon $\omega$, which has spectrum $d\omega/\omega$.
This corresponds to three pomerons term in inclusive transverse
cross section
$\displaystyle\frac{d^2\sigma}{dtdM^2}\sim\frac{1}{1-x}$.

A qualitative result comes after this assumption. Multiplicity
distribution in diffractive beam will be different from
multiplicity distributions in $pp$ and $p\bar{p}$ interactions.

Quite evidently that contribution from diagram in  Fig.~7b equals
to contribution from diagram in  Fig.~7c multiplied by coefficient
of shower enhancement proposed by Kaidalov. Moreover, since only
one gluon from eight in Fig.~7a gives colorless quark system, then
contribution from diagram in Fig.~7c equals to one twelfth
contribution from diagram in Fig.~7a. The same arguments are
correct to configuration with two additional gluons.

We estimate the upper limit of $\sigma^{sd}$ by taking one twelfth
from contributions to total cross sections of configurations with
one and two additional gluons and variable constant term (Fig.~8,
data were taken from~\cite{bib7}).

$$\begin{array}{l}
\displaystyle\sigma^{sd}=\sigma_0^{sd}+\frac{1}{12}\,\sigma_1^{tot}\ln
s+\frac{1}{12}\,\sigma_2^{tot}(\ln
s)^2,\\[3mm]
\displaystyle\sigma_0^{sd}\simeq4.2\;\mbox{mb},\;\;\sigma^{sd}\simeq11\;\mbox{mb
at 14 TeV}
\end{array}
$$

Configuration with one additional gluon corresponds to super hard
component $\delta(1-z)$ in pomeron structure function, since all
longitudinal momentum that is used for diffractive beam production
is carried by gluon; $z$ - part of pomeron momentum carried by
gluon.

Configuration with two additional gluons corresponds to hard
component $z(1-z)$.

\end{document}